\documentclass[aps,prd,showkeys,showpacs,amssymb,
twocolumn,
amsfonts,epsf,preprintnumbers,nofootinbib,superscriptaddress]{revtex4}

\usepackage[dvips]{graphicx}
\usepackage{bm,latexsym,amsmath,amssymb,amsfonts,color}
\newcommand\beq{\begin{eqnarray}}
\newcommand\eeq{\end{eqnarray}}

\begin{document}

\title{Self-accelerating solutions in the cascading DGP braneworld}

\author{Masato Minamitsuji}
\email{minamituzi_at_sogang.ac.kr}
\affiliation{Center for Quantum Spacetime, Sogang University,
Shinsu-dong 1, Mapo-gu, Seoul, 121-742 South Korea}

\begin{abstract}
The self-accelerating branch of
the Dvali-Gabadadze-Porrati (DGP) five-dimensional braneworld has provided
a compelling model for the current cosmic acceleration.
Recent observations, however, have not favored it so much.
We discuss the solutions which contain
a de Sitter 3-brane in the cascading DGP braneworld model,
which is a kind of higher-dimensional generalizations of the DGP model,
where a $p$-dimensional brane is placed on a $(p+1)$-dimensional one
and the $p$-brane action contains the $(p+1)$-dimensional induced
scalar curvature term.
In the simplest six-dimensional model,
we derive the solutions.
Our solutions can be classified into two branches,
which reduce to the self-accelerating and normal solutions
in the limit of the original five-dimensional DGP model.
In the presence of the six-dimensional bulk gravity,
the `normal' branch provides a new self-accelerating solution.
The expansion rate of this new branch is
generically lower than that of the original one,
which may alleviate the fine-tuning problem.
\end{abstract}

\maketitle

Recent observational data with high precision suggest that our
Universe is currently in an accelerating phase \cite{1,2}.
They are
consistent
with the presence of a nonzero cosmological constant or quantum vacuum
energy, but its value must be extremely tiny.
In the context of the braneworld, the
Dvali-Gabadadze-Porrati (DGP) five-dimensional model
has been a compelling model for the cosmic acceleration \cite{3,4,5}.
The DGP model contains a
mechanism to modify the gravitational law just on cosmological scales by
the effects of the four-dimensional Einstein-Hilbert term put
into the action of our 3-brane Universe.
Such an intrinsic curvature term would be induced
due to quantum loops of the matter fields which are localized on the 3-brane.
The effect of the
four-dimensional intrinsic curvature term on the 3-brane recovers
the Einstein gravity on small scales but on large distance
scales gravitational law becomes five-dimensional.
The DGP model realizes
the so-called self-accelerating Universe that features a four-dimensional de
Sitter phase even though our 3-brane Universe is completely empty.
Recent studies, however,
have indicated that the observational data have not
favored the self-accelerating branch of DGP \cite{obs_dgp}.
The self-accelerating solutions
have also faced the disastrous issue of ghost excitations \cite{8}:
The energy is not bounded from below and therefore
the theory is already pathological
even at the classical level.

There are possibilities that the realistic cosmological model may be
obtained by generalizing the five-dimensional DGP model to a
higher-dimensional spacetime.
An extension of the DGP model,
where two 4-branes with the induced gravity terms are
intersecting in the six-dimensional spacetime, has been
investigated in Ref. \cite{9}.
Another interesting model is so-called the
cascading DGP model \cite{6}.
The model is constructed by a set of branes
of the different dimensionality,
where a $p$-brane is placed on a $(p+1)$-dimensional brane and
the $p$-brane action contains the
$(p+1)$-dimensional induced scalar curvature term.
For instance, in the simplest six-dimensional model,
a 3-brane Universe whose action contains an induced four-dimensional scalar
curvature term
is placed on a 4-brane
whose action contains an induced five-dimensional scalar
curvature term,
embedded into a (possibly infinitely extended) six-dimensional spacetime.
An extension to the case of an arbitrary number of spacetime dimensions
is straightforward in principle.
It is expected that in such kind of model,
in the infrared region the gravitational force falls off
sufficiently fast to exhibit `degravitation' \cite{6}.
The linearized analysis has confirmed this idea in part
at the level of the linearized theory \cite{6}.

One of crucial questions is the viability of the cascading DGP model.
To answer to this question, of course, one should go beyond
the linearized analysis and in particular investigate the cosmology.
Non-linearities may detect effects which
may not appear in the linearized treatment.
In addition, cosmology can
help to have a better understanding of the model and of the idea of
gravity localized through intrinsic curvature terms
on the 3-brane and 4-brane.
As the first step to this direction,
we will look for the solutions which contain a de Sitter 3-brane.
They may give rise to the self-accelerating cosmological solutions
in the simplest six-dimensional cascading DGP model.

%%%%%%%%%%%%%%%%%%%%%%%%%%%%%%%%%%%%%%%%%

\vspace{0.1cm}

The system of our interest is that our 3-brane Universe
$\Sigma_4$ is placed on a 4-brane $\Sigma_5$, embedded into the
six-dimensional bulk ${\cal M}_6$.
For simplicity, we suppress the matter terms in the bulk
and on the branes.
The total action is given by
 \beq
\label{action}
S&=&
\frac{M_6^4}{2}\int_{{\cal M}_6} d^6X \sqrt{-G}{}^{(6)}{R}
+\frac{M_5^3}{2}
\int_{\Sigma_5} d^5 y \sqrt{-q}{}^{(5)}{R}
\nonumber\\
&+&\frac{M_4^2}{2}
\int_{\Sigma_4} d^4 x\sqrt{-g} {}^{(4)}{R},
\eeq
where $G_{AB}$,
$q_{ab}$ and $g_{\mu\nu}$ represent metrics in ${\cal M}_6$,
on $\Sigma_5$ and $\Sigma_4$, respectively.
${}^{(i)}R$ ($i=6,5,4$) are
Ricci scalar curvature terms
associated with respect to $G_{AB}$, $q_{ab}$ and $g_{\mu\nu}$.
For the later discussion, it is useful to
introduce the crossover mass scales $m_5:= M_5^3/M_4^2$ and $m_6:=M_6^4/M_5^3$,
which determines the energy scale where the five-dimensional
and six-dimensional physics appear, respectively.
We assume that $m_5>m_6$.
Then, it is natural to expect that
the effective gravitational theory becomes
four-dimensional for $H> m_5$,
five-dimensional for $m_5> H > m_6$,
and finally six-dimensional for $H< m_6$,
where $H$ is the cosmic expansion rate.

We consider the six-dimensional Minkowski spacetime, which is
covered by the following choice of the coordinates
\beq
ds_6^2=G_{AB}dX^A dX^B=dr^2+d\theta^2+H^2 r^2\gamma_{\mu\nu}dx^{\mu}dx^{\nu},
\label{6d}
\eeq
where $\gamma_{\mu\nu}$ is the metric of the four-dimensional
de Sitter spacetime with the expansion rate $H$.
The $r$ and $\theta$ coordinates represent two extra dimensions and
$x^{\mu}$ ($\mu=0,1,2,3$) do the ordinary four-dimensional spacetime.
%%%%%%%%%%%%%%%%%%%%%%%%%%%%%%%%%%%%%%%%%%%%%%%%%%%
The surface of $r=0$ corresponds to
a (Rindler-like) horizon and 
only the region of $r\geq 0$ is considered.
Note that the boundary surface of $r=0$ 
does not cause any pathological effect
because it is not a singularity.
%%%%%%%%%%%%%%%%%%%%%%%%%%%%%%%%%%%%%%%%%%%%%%%%%%%
We consider a 4-brane located along the trajectory
$(r(|\xi|),\theta(|\xi|))$, where
the affine parameter $\xi$ gives the proper coordinate
along the 4-brane.
The 3-brane is placed at $\xi=0$,
and for decreasing value of $|\xi|$ one approaches the 3-brane.
We assume the $Z_2$-symmetry across the 4-brane and hence
an identical copy is glued to the opposite side.
Along the trajectory of the 4-brane
$\dot{r}^2+\dot{\theta}^2=1$,
where the dot represents the derivative with respect to $\xi$.
The induced metric on the 4-brane is given by
\beq
ds_5^2=q_{ab}dy^a dy^b
=d\xi^2+H^2r(|\xi|)^2\gamma_{\mu\nu}dx^{\mu}dx^{\nu}.
\eeq
%%%%%%%%%%%%%%%%%%%%%%%%%%%%%%%%%%%%%%%%%%%
The point where $r(\xi)=0$ on the 4-brane
corresponds to a horizon
and the 4-brane is not extended beyond it.
%%%%%%%%%%%%%%%%%%%%%%%%%%%%%%%%%%%%%%%%%%%%%%%%%%%%%%%%
The 3-brane geometry is exactly de Sitter spacetime
with the normalization condition $Hr(0)=1$,
\beq
\label{4d}
ds_4^2=g_{\mu\nu}dx^{\mu}dx^{\nu}
=\gamma_{\mu\nu}dx^{\mu}dx^{\nu}.
\eeq
The nonvanishing
components of the tangential and normal vectors to the 4-brane are given by
\beq
u^r=\dot{r},\quad
u^{\theta}=\dot{\theta},\quad
n^r=\epsilon\dot{\theta},\quad
n^{\theta}=-\epsilon \dot{r}.
\label{comp}
\eeq
We restrict that the region to be considered is to be $r>0$
and the 3-brane is sitting on $r$-axis ($\theta=0$).
The 4-brane trajectory is $Z_2$-symmetric
across the 3-brane.
$\dot{\theta}>0$ for increasing $\xi$.
In the case of $\epsilon=1$,
the bulk space is in the side of increasing $r$,
while
in the case of $\epsilon=-1$,
the bulk space is the side of decreasing $r$.

The components of
the extrinsic curvature tensor defined by
$K_{ab}:=\nabla_a n_b$
are given by
\beq
&& K_{\xi\xi}-K
=-\frac{4\epsilon}{r}\big(1-\dot{r}^2\big)^{1/2},
\nonumber\\
&& K_{\mu\nu}-q_{\mu\nu}K
=\epsilon
\Big(-\frac{3}{r}\big(1-\dot{r}^2\big)^{1/2}
+\frac{\ddot{r}}{(1-\dot{r}^2)^{1/2}}
\Big)q_{\mu\nu}.
\eeq
The junction condition is given by
\beq
M_6^4\Big[ K_{ab}-q_{ab}K\Big]=
\left(M_5^3 {}^{(5)}G_{ab}
+M_4^2{}^{(4)}G_{\mu\nu}\delta^{\mu}_a\delta^{\nu}_b\delta(\xi)
\right).
\eeq
where the square bracket denotes the jump of a bulk quantity across the 4-brane
and
the components of the Einstein tensor on the 4- and 3-branes are given by
\beq
&&{}^{(5)}G_{\xi\xi}
=-6\frac{1-\dot{r}^2}{r^2},
\nonumber\\
&&{}^{(5)}G_{\mu\nu}
=3\frac{\dot{r}^2+r\ddot{r}-1}{r^2}q_{\mu\nu},
\eeq
and ${}^{(4)}G_{\mu\nu}=-3H^2\gamma_{\mu\nu}$.
By taking the $Z_2$-symmetry across the 4-brane
into consideration,
the matching condition becomes
\beq\label{y2}
&&
-M_6^4\epsilon \frac{4}{r}\big(1-\dot{r}^2\big)^{1/2}
=-3M_5^3\frac{1-\dot{r}^2}{r^2},
\nonumber\\
&&
M_6^4\epsilon
\Big(-\frac{3}{r}
 \big(1-\dot{r}^2\big)^{1/2}
 +\frac{\ddot{r}}{(1-\dot{r}^2)^{1/2}}
\Big)
\nonumber\\
&=&\frac{3M_5^3}{2}
\frac{\dot{r}^2+r\ddot{r}-1}
     {r^2}.
\eeq
The way to construct the solution is essentially the same as the case of
a tensional 3-brane on a tensional 4-brane (See Appendix).

In our case, it is suitable to take $\epsilon=+1$ branch.
Then, the junction condition tells that
the trajectory of the 4-brane is given by
$r(\xi)=a^{-1}\cos(a|\xi|-a\xi_0)$ with
\beq
a=\frac{4m_6}{3}. \label{a}
\eeq
where we assume $0<a\xi_0<\pi/2$.
%%%%%%%%%%%%%%%%%%%%%%%%%%%%%%%%%%%%%%%
$r(\xi)$ vanishes at $|\xi|=|\xi_{\rm max}|=\pi/(2a)+\xi_0$.
Note that, as mentioned before,
the surface of $r=0$
corresponds to a horizon
and on the 4-brane there are horizons at $|\xi|=|\xi_{\rm max}|$, 
namely at a finite proper distance from the 3-brane. 
The 4-brane is not extended beyond them \cite{dw}.
%%%%%%%%%%%%%%%%%%%%%%%%%%%%%%%%%%%%%%%%%%%%
Now an identical copy is attached across the 4-brane.
The normalization of the overall factor of the metric function at the 3-brane
place requires $\cos(a\xi_0)=a/H\leq 1$.
Note that
\beq
H\geq \frac{4m_6}{3}.
\eeq
The $\ddot{r}$ term gives rise to the contribution
proportional to $\delta(\xi)$.
Here, by noting that
\beq
\frac{d}{d\xi}{\rm arctan}
\left(\frac{\dot{r}}{\sqrt{1-\dot{r}^2}}\right)
=\frac{\ddot{r}}{\sqrt{1-\dot{r}^2}},
\eeq
and integrating the $(\mu,\nu)$-component of
the junction equation Eq. (\ref{y2}) across $\xi=0$,
one finds
\beq
M_6^4(4a\xi_0)
=6M_5^3 a\tan(a\xi_0)-3H^2 M_4^2,
\eeq
which with Eq. (\ref{a}) leads to
\beq
\label{H2}
\frac{H}{2m_5}
-\Big(
\sqrt{1-\frac{16m_6^2}{9H^2}}
-\frac{2m_6}{3H}
 {\rm arctan}
\big(
\sqrt{\frac{9H^2}{16m_6^2}-1}\big)\Big)
=0\,.
\eeq
The solution of Eq. (\ref{H2}) determines the value of
the expansion rate $H$.
The 3-brane induces the deficit angle $4a\xi_0$ in the bulk.
The configuration of the bulk space is shown in Fig. 1.
%%%%%%%%%%%%%%%%%%%%%%%%%%%%%%%%%%%%
The bulk space is outside the curve of the 4-brane
and has an infinite volume.
As mentioned before,
the surface of $r=0$ corresponds to
a horizon and, in particular,
on the 4-brane there are horizons
at a finite proper distance from the 3-brane.
The 4-brane is not extended beyond them.
Note that this surface does not cause
any pathological effect.
%%%%%%%%%%%%%%%%%%%%%%%%%%%%%%%%%%%%%%%%%%%%%%
\begin{figure}
\begin{minipage}[t]{.45\textwidth}
\label{fig5}
   \begin{center}
    \includegraphics[scale=.50]{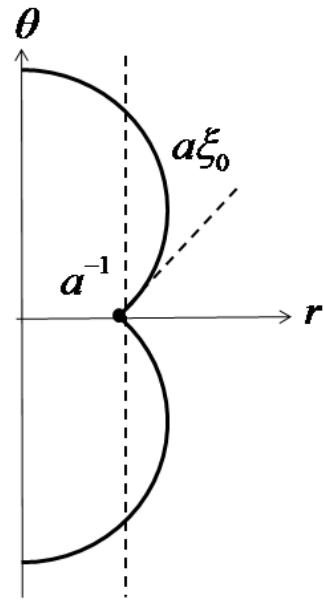}
\vspace{-.10cm}
        \caption{
%%%%%%%%%%%%%% new %%%%%%%%%%%%%%%%%%%%%%%%%%%%
The configuration of the bulk space is shown.
The circle point and solid curve represent
the 3- and 4-branes, respectively.
According to the direction of the normal vector,
the bulk space is outside the curve of the 4-brane.
In this picture,
each point represents the four-dimensional de Sitter
spacetime.
Because of the $Z_2$-symmetry, an identical copy
of this picture is glued across the 4-brane.
The bulk space is outside the curve of the 4-brane
and has an infinite volume.
The surface of $r=0$ corresponds to
a horizon and, in particular,
on the 4-brane there are horizons
at a finite proper distance from the 3-brane.
The 4-brane is not extended beyond them.
%%%%%%%%%%%%%%%%%%%%%%%%%%%%%%%%%%%%%%%%%%%%%%%
}
   \end{center}
 \end{minipage}
\end{figure}
%%%%%%%%%%%%%%%%%%%%%%%%%%%%%%%%%%%%%%%%%%%%%%

For generic values of $m_6$,
in Fig 1, the left-hand-side of Eq. (\ref{H2})
is shown as a function of $H/m_5$ for each fixed ratio $m_6/m_5$.
%%%%%%%%%%%%%%%%%%%%%%%%%%%%%%%%%%%%%%%%%%%%%%
\begin{figure}
\begin{minipage}[t]{.45\textwidth}
\label{fig1}
   \begin{center}
    \includegraphics[scale=.80]{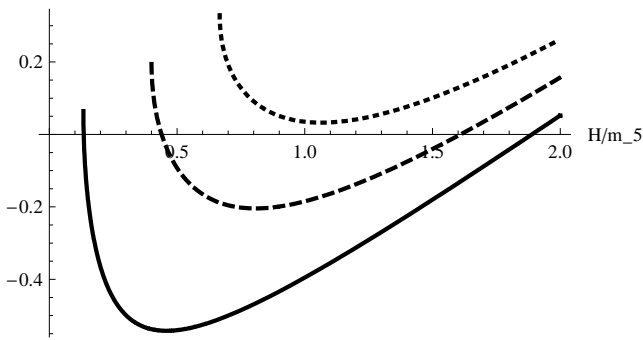}
\vspace{.0cm}
        \caption{
The left-hand-side of Eq. (\ref{H2})
is shown as a function of $H/m_5$ for fixed ratio $m_6/m_5$.
The solid, dashed and dotted curves correspond to the cases of
$m_6/m_5=0.1,0.3,0.5$, respectively.
In the last case,
in which $m_6/m_5$
is above the critical value $\big(m_6/m_5\big)_{\rm crit}=0.46978$,
there is no solution.
}
   \end{center}
 \end{minipage}
\end{figure}
%%%%%%%%%%%%%%%%%%%%%%%%%%%%%%%%%%%%%%
It is found that below the critical ratio
$m_6/m_5<\big(m_6/m_5\big)_{\rm crit}\approx 0.46978$,
there are two branches of solutions, which are here denoted
by $H_+>H_-$.
On the other hand, for $m_6/m_5>\big(m_6/m_5\big)_{\rm crit}$,
there is no solution of Eq. (\ref{H2}).
In the marginal case of $m_6/m_5=\big(m_6/m_5\big)_{\rm crit}$,
there is the degenerate solution given by $H\approx m_5$.
For generic values of $m_6/m_5$,
in Fig. 2 and 3, the solutions $H_+$ and $H_-$ are shown as
functions of $m_6/m_5(<\big(m_6/m_5\big)_{\rm crit})$, respectively.
%%%%%%%%%%%%%%%%%%%%%%%%%%%%%%%%%%%%%%%%%%%%%%
\begin{figure}
\begin{minipage}[t]{.45\textwidth}
\label{fig2}
   \begin{center}
    \includegraphics[scale=.80]{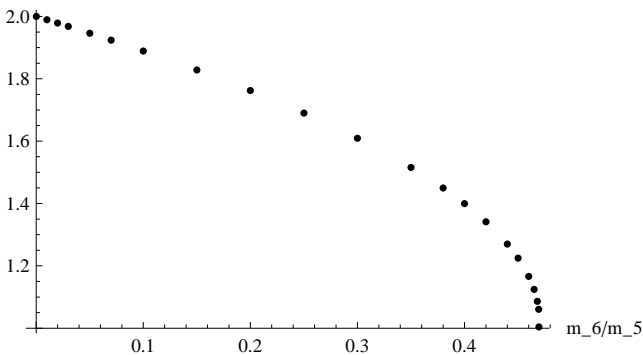}
        \caption{
The larger solution $H_+$ is shown as a function of $m_6/m_5$,
in the unit of $m_5$.
}
   \end{center}
 \end{minipage}
\end{figure}
\begin{figure}
\begin{minipage}[t]{.45\textwidth}
\label{fig2}
   \begin{center}
    \includegraphics[scale=.80]{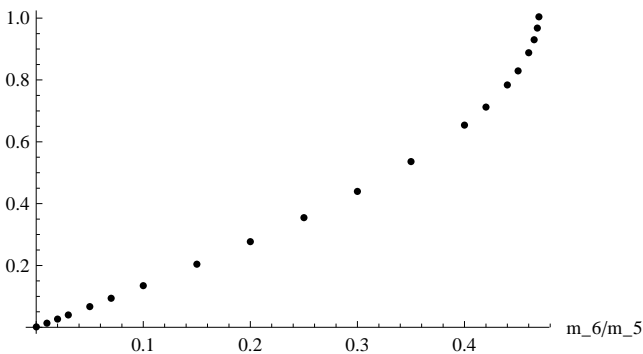}
        \caption{
The smaller solution $H_-$ is shown as a function of $m_6/m_5$,
in the unit of $m_5$.
}
   \end{center}
 \end{minipage}
\end{figure}
%%%%%%%%%%%%%%%%%%%%%%%%%%%%%%%%%%%%%%%%%%
In the limit of $m_6\ll m_5$, another solution
is given approximately given by
\beq
H_+\approx 2m_5,\quad
H_-\approx \frac{4m_6}{3}.
\eeq
In the absence of the bulk gravity, $m_6\to 0$,
the $(+)$ and $(-)$-branches
coincide with the `self-accelerating'
and `normal' solutions in the DGP model,
with $H_+=2m_5$ and $H_-=0$, respectively.
By taking the presence of the six-dimensional bulk into consideration,
the self-accelerating branch essentially remains the same.
But the normal branch solution provides
a new self-accelerating solution if $H_-$,
which could be much smaller than $H_+$ for $m_6\ll m_5$.
Note that the existence of both of these new solutions relies on the presence of the 4-brane,
since in the limit of $M_5\to 0$ none of these solutions can exist.

As we mentioned,
the self-accelerating branch of the original DGP model
is not favored by recent observations and
also suffers a ghost instability.
What we found is that in the six-dimensional cascading DGP model,
one of two branches,
which corresponds to the `normal' branch in the
original DGP model,
provides a new self-accelerating solution
whose expansion rate could be much smaller than
that in the other branch,
which corresponds to the original `self-accelerating' branch.
Thus, the fine-tuning would be relaxed in some degrees.
%%%%%%%%%%%%%%%%%% new %%%%%%%%%%%%%%%%%%%%%%%%%%%%
In the self-accelerating solution of the DGP model,
the bulk spacetime is infinitely extended
and a mode which satisfies the background solution is not normalizable.
Thus,
the scalar mode is hence different from the zero mode,
which already implies the potential pathology about the ghost instability.
In our new solutions
the 4-brane where the 3-brane resides
can never reach the infinity (see Fig. 1) and has a finite volume.
Therefore, in analogy with the case of the standard DGP,
it implies that the bending mode of the 3-brane would be normalized
and hence solutions could be healthy,
although the detailed investigations about the stability
are left for a future work.
%%%%%%%%%%%%%%%%%%%%%%%%%%%%%%%%%%%%%%%%%%%%%%%

The idea of the cascading gravity may be extendable
to the case of an arbitrary number of spacetime dimensions.
In the case of $n$-dimensional spacetime,
there would be 3-, 4-,$\cdots$, $(n-2)$-branes, where
a $p$-brane ($p=3,4\cdots,n-3$) is placed on a $(p+1)$-dimensional brane
and in the $p$-brane action
the $(p+1)$-dimensional scalar curvature term
is induced with the coupling constant $1/M_{p+1}^{p-1}$.
Assuming the hierarchal relation among the crossover scales
$m_n\ll m_{n-1}\ll m_{n-2}\ll\cdots\ll m_{6}\ll m_5$,
where $m_q=M_{q}^{q-2}/M_{q-1}^{q-3}$ ($q=5,6,\cdots,n$),
the solution with the smallest expansion rate
would be given by $H\simeq m_n$.
Thus, the resultant expansion rate becomes tiny and
the presence of enough number of branes
may resolve the fine-tuning problem.
%This may support the idea of degravitation,
%which is expected in terms of the linearized analysis \cite{6}.

\section*{Acknowledgement}
The author wishes to thank the anonymous reviewers for his/her comments.
This work was supported by the National Research Foundation of Korea (NRF) grant funded by the Korea government(MEST) (No. 20090063070).

\appendix

\section{The case of pure tension branes}

In the system composed of tensional 3- and 4-branes
in the six-dimensional bulk, the action is given by
\beq
\label{action}
S&=&
\frac{M_6^4}{2}\int_{{\cal M}_6} d^6X \sqrt{-G}{}^{(6)}{R}
+
\int_{\Sigma_5} d^5 y \sqrt{-q}\big(-\sigma_4\big)
\nonumber\\
&+&
\int_{\Sigma_4} d^4 x\sqrt{-g}\big(-\sigma_3\big),
\eeq
where $\sigma_3$ and $\sigma_4$ are tensions of branes.
We assume that both the brane tensions are positive.
Then, we look for the de Sitter 3-brane solution.

The ansatz of the spacetime metric is assumed to be the same
as the case in the text, discussed in Eq (\ref{6d})-(\ref{4d}).
Then, the junction condition becomes
\beq
M_6^4\Big[ K_{ab}-q_{ab}K\Big]=
 \sigma_4 q_{ab}
+\sigma_3 g_{\mu\nu}\delta^{\mu}_a\delta^{\nu}_b\delta(\xi).
\eeq

The matching conditions on the 4-brane are given by
\beq
&&
-M_6^4\epsilon \frac{4}{r}\big(1-\dot{r}^2\big)^{1/2}
=\frac{\sigma_4}{2},
\nonumber\\
&&
M_6^4\epsilon
\Big(-\frac{3}{r}
 \big(1-\dot{r}^2\big)^{1/2}
 +\frac{\ddot{r}}{(1-\dot{r}^2)^{1/2}}
\Big)
=\frac{\sigma_4}{2}.
\eeq
To obtain a 4-brane with a positive tension,
it is suitable to choose $\epsilon=-1$ branch.
The trajectory is  given by
$r(\xi)=a^{-1}\cos(a|\xi|+a\xi_1)$ with
\beq
\sigma_4=8M_6^4 a
\eeq
where we assume $0<a\xi_1<\pi/2$.
%%%%%%%%%%%%%%%%%%%%%%%%%%%%%%%%%%%%%%%%%%%%%%%
$r(\xi)$ vanishes at $|\xi|=|\xi_{\rm max}|=\pi/(2a)-\xi_1$.
Note that
the surface of $r=0$
corresponds to a horizon
and on the 4-brane there are horizons at $|\xi|=|\xi_{\rm max}|$, 
namely at a finite proper distance from the 3-brane. 
The 4-brane is not extended beyond them \cite{dw}.
%%%%%%%%%%%%%%%%%%%%%%%%%%%%%%%%%%%%%%%%%%%%%%%%%
On the other hand, the 3-brane induces the deficit angle
given by $4a\xi_1$, which is determined through the 3-brane junction condition
as
\beq
\sigma_3=M_6^4(4a\xi_1),
\eeq
which is the standard tension-deficit relation for a conical singularity.
The normalization condition of 3-brane metric provides
$\cos(a\xi_1)=a/H$.
It gives the expansion rate $H$ in terms of the 3-brane and 4-brane tension as
\beq
H=\frac{\sigma_4}{8M_6^4\cos\big(\frac{\sigma_3}{4M_6^4}\big)},
\eeq
which is the six-dimensional generalization of the case
without the bulk cosmological constant,
discussed in Ref. \cite{dw}.
%%%%%%%%%%%%%%%%%%% new %%%%%%%%%%%%%%%%%%%%%%%%%%%%%%%%
The configuration of the bulk space is shown in Fig. 5.
The bulk space is inside the curve of the 4-brane
and has a finite volume.
As mentioned before,
the surface of $r=0$ corresponds to
a horizon and, in particular,
on the 4-brane there are horizons
at a finite proper distance from the 3-brane.
The 4-brane is not extended beyond them.
Note that this surface does not cause
any pathological effect.
%%%%%%%%%%%%%%%%%%%%%%%%%%%%%%%%%%%%%%%%%%%%%%%%%%
%%%%%%%%%%%%%%%%%%%%%%%%%%%%%%%%%%%%%%%%%%%%%%
\begin{figure}
\begin{minipage}[t]{.45\textwidth}
\label{fig6}
   \begin{center}
    \includegraphics[scale=.50]{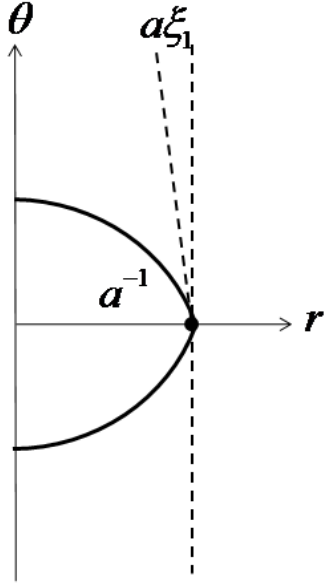}
\vspace{-.10cm}
        \caption{
%%%%%%%%%%%%%%%%%%%%%%%%  new %%%%%%%%%%%%%%%%%%%%%%%%%%%
The configuration of the bulk space is shown.
The circle point and solid curve represent
the 3- and 4-branes, respectively.
According to the direction of the normal vector,
the bulk space is inside the curve of the 4-brane.
In this picture,
each point represents the four-dimensional de Sitter
spacetime.
Because of the $Z_2$-symmetry, an identical copy
of this picture is glued across the 4-brane.
The bulk space is inside the curve of the 4-brane
and has a finite volume.
The surface of $r=0$ corresponds to
a horizon and, in particular,
on the 4-brane there are horizons
at a finite proper distance from the 3-brane.
The 4-brane is not extended beyond them.
%%%%%%%%%%%%%%%%%%%%%%%%%%%%%%%%%%%%%%%%%%%%%%%%%%%
}
   \end{center}
 \end{minipage}
\end{figure}
%%%%%%%%%%%%%%%%%%%%%%%%%%%%%%%%%%%%%%%%%%%%%%

\end{document}